 \documentclass[prb,twocolumn,showpacs,showkeys,superscriptaddress]{revtex4}

\usepackage{graphicx}
\usepackage{amssymb,amsmath}
\usepackage{dcolumn}
\usepackage{bm}
\usepackage{color}
\usepackage{tabularx}

\begin{document}

\title{On the strong impact of doping in the triangular antiferromagnet 
       $ {\bf CuCrO_2}$} 

\author{Antoine Maignan}
\affiliation{Laboratoire CRISMAT, UMR CNRS-ENSICAEN(ISMRA) 6508, 
             and IRMA, FR3095, Caen, France}
\author{Christine Martin}
\affiliation{Laboratoire CRISMAT, UMR CNRS-ENSICAEN(ISMRA) 6508, 
             and IRMA, FR3095, Caen, France}
\author{Raymond   Fr\'esard}
\affiliation{Laboratoire CRISMAT, UMR CNRS-ENSICAEN(ISMRA) 6508, 
             and IRMA, FR3095, Caen, France}
\author{Volker Eyert}
\affiliation{Laboratoire CRISMAT, UMR CNRS-ENSICAEN(ISMRA) 6508, 
             and IRMA, FR3095, Caen, France}
\affiliation{Center for Electronic Correlations and Magnetism, 
             Institut f\"ur Physik, Universit\"at Augsburg, 
             86135 Augsburg, Germany}
\author{Emmanuel Guilmeau}
\affiliation{Laboratoire CRISMAT, UMR CNRS-ENSICAEN(ISMRA) 6508, 
             and IRMA, FR3095, Caen, France}
\author{Sylvie H\'ebert}
\affiliation{Laboratoire CRISMAT, UMR CNRS-ENSICAEN(ISMRA) 6508, 
             and IRMA, FR3095, Caen, France}
\author{Maria Poienar}
\affiliation{Laboratoire CRISMAT, UMR CNRS-ENSICAEN(ISMRA) 6508, 
             and IRMA, FR3095, Caen, France}
\author{Denis Pelloquin}
\affiliation{Laboratoire CRISMAT, UMR CNRS-ENSICAEN(ISMRA) 6508, 
             and IRMA, FR3095, Caen, France}

\date{\today}

\begin{abstract}
Electronic band structure calculations using the augmented spherical wave
method have been performed for $ {\rm CuCrO_2}$. For this antiferromagnetic
($T_N$ = 24 K) semiconductor crystallizing in the delafossite structure, it is
found that the valence band maximum is mainly due to the t$_{2g}$ orbitals of
Cr$^{3+}$ and that spin polarization is predicted with 3 $\mu_B$ per
Cr$^{3+}$. The structural characterizations of $ {\rm CuCr_{1-x} Mg_x O_2}$
reveal a very limited range of Mg$^{2+}$ substitution for Cr$^{3+}$ in this
series. As soon as $ {\rm x} = 0.02$, a maximum of 1\% Cr ions substituted by Mg
site is measured in the sample. This result is also consistent with the
detection of Mg spinel impurities from X-ray diffraction for x = 0.01. This
explains the saturation of the Mg$^{2+}$ effect upon the electrical
resistivity and thermoelectric power observed for $ {\rm x} > 0.01$. Such a very weak
solubility limit could also be responsible for the discrepancies found in the
literature. Furthermore, the measurements made under magnetic field (magnetic
susceptibility, electrical resistivity and Seebeck coefficient) support that
the Cr$^{4+}$ "‘holes"’, created by the Mg$^{2+}$ substitution, in the matrix
of high spin Cr$^{3+}$ (S = 3/2) are responsible for the transport properties
of these compounds. 
\end{abstract}


\maketitle

\section{Introduction}
\label{sec:intro}
Because of their remarkable properties, layered oxides with mixed-valent
transition-metal atoms have attracted much attention in the last twenty
years. After the discovery of the high-T$_c$ superconducting cuprates
[\onlinecite{Bednorz}]  with
two-dimensional (2D) structures, in which perovskite $[{\rm ACuO_{3-\delta}
}]_{\infty}$ layers are sandwiched in between layers acting as charge
reservoirs, the layered cobaltites based on ${\rm CoO_2}$ layers with the
${\rm CdI_2}$ structure have been more recently deeply investigated. Apart
from the cationic conduction in ${\rm Li_x CoO_2}$, an effective battery
material [\onlinecite{Mitzushima}], the analogous compound ${\rm Na_x CoO_2}$
has been shown to 
exhibit promising thermoelectric properties for $ {\rm x} \sim 0.5$
[\onlinecite{Terasaki}]  whereas the
hydrated form of ${\rm Na_{0.3} CoO_2}$ exhibits superconductivity [\onlinecite{Takada}]. From
the structural point of view, the ${\rm CoO_2}$ layers can be described as
planes of edge sharing ${\rm CoO_6}$ octahedra forming a 2D triangular cobalt
lattice [\onlinecite{Jansen}]. This triangular network is responsible for the frustrated
magnetism. The particular rhombohedral splitting of the cobalt $t_{2g}$
orbitals has been invoked to explain the coexistence of two subbands, $a_{1g}$
and $e'_g$, narrow and broad, respectively, the former being responsible for
the large Seebeck coefficient whereas the metallic behaviour is associated 
with the latter [\onlinecite{Singh00}].

In this context, the coexistence of complex magnetic structure
[\onlinecite{Takeda}] and ferroelectricity [\onlinecite{Kimura06}] in the
delafossite structure ${\rm  CuFeO_2}$ 
is very interesting. In this compound crystallizing in the $R\bar{3}m$ space
group, the ${\rm FeO_6}$ octahedra layers are isostructural to the ${\rm CoO_6}$ layer
of the ${\rm Na_x CoO_2}$ system [\onlinecite{Pabst}]. The main difference lies in the separating layer with a linear coordination of monovalent copper,
since the O--Cu--O bridges ensure the connection between successive ${\rm
  FeO_2}$ planes. The presence of high spin (S = 5/2) at the 
B-site of the ${\rm AFeO_2}$ delafossite together with the triangular spin
frustration creates complex antiferromagnetic states, coupled to
structural transitions, with modulations of the magnetic structure linked to ferroelectricity [\onlinecite{Kimura06,Ye}]. However, the very stable high
spin configuration of ${\rm Fe^{3+}}$, precluding the possibility to induce
electrical conductivity, is in marked contrast with the low electrical resistivity reported for the chromium based analogous delafossite, ${\rm CuCr_{1-x} Mg_x O_2}$ [\onlinecite{Okuda,Zcucrmg}]. The ${\rm CuCrO_2}$ compound is clearly
insulating and its magnetic structure, established by neutron
diffraction data refinements, revealed a complex antiferromagnetic incommensurate structure with a short magnetic correlation
length along c and a maximal magnetic moment of $ \sim 3 \mu_B$
[\onlinecite{Kadowaki,Poienar}]. 
Clear evidence for magnetic ordering below $T_N$ is provided by specific heat measurements as well, for both undoped and Mg-doped
samples [\onlinecite{Okuda}], the former being also a multiferroic
[\onlinecite{Seki08}]. However, 
the electrical resistivity drop induced by the ${\rm Mg^{2+}}$ for ${\rm Cr^{3+}}$ substitution and its relation to the structural properties has not yet
been clearly elucidated. The values of the transport coefficients
show discrepancies, too. Indeed, according to Ono et
al. [\onlinecite{Zcucrmg}], the 
resistivity of ${\rm CuCr_{0.98} Mg_{0.02} O_2}$ at room temperature (RT) is
about $1~{\rm \Omega~cm}$, while according to Okuda et al. it is rather one order of
magnitude smaller [\onlinecite{Okuda}]. Regarding the thermopower at room temperature, the former authors reported a value around $280 \mu {\rm V/K}$
whereas the latter obtained $100 \mu {\rm V/K}$. Besides, some authors
mentioned a hole creation at the A site by charge compensation,
${\rm Cu^+_{1-x} Cu^{2+}_x Cr^{3+}_{1-x} Mg^{2+}_x O_2}$ [\onlinecite{Okuda}],
whereas others considered a mixed valency ${\rm Cr^{3+}/Cr^{4+}}$ according
to ${\rm Cu^+ Cr^{3+}_{1-2x} Cr^{4+}_{x} Mg^{2+}_x O_2}$
[\onlinecite{Zcucrmg}]. For all these reasons  
the ${\rm CuCr_{1-x} Mg_x O_2}$ system has been revisited.

In the present paper, we report on electronic band structure
calculations for the undoped compound ${\rm CuCrO_2}$ and on a complete
experimental study of the series ${\rm CuCr_{1-x} Mg_x O_2}$. Special care has
been taken with the chemistry problem of the solubility of ${\rm Mg^{2+}}$
at the chromium site.

\section{Methodology}
\subsection{Electronic structure calculations: Theoretical method}

The electronic band structure calculations were based on
density-functional theory and the generalized gradient approximation (GGA)
[\onlinecite{Perdew96}] with the local-density approximation parameterized
according to Perdew and Wang [\onlinecite{Perdew92}]. They were 
performed using the scalar-relativistic implementation of the
augmented spherical wave (ASW) method (see
Refs. [\onlinecite{wkg,aswrev,aswbook}] and 
references therein). In the ASW method, the wave function is
expanded in atom-centered augmented spherical waves, which are
Hankel functions and numerical solutions of Schr\"odinger’s equation,
respectively, outside and inside the so-called augmentation 
spheres. In order to optimize the basis set, additional augmented
spherical waves were placed at carefully selected interstitial sites.
The choice of these sites as well as the augmentation radii were
automatically determined using the sphere-geometry optimization algorithm
[\onlinecite{sgo}]. Self-consistency was achieved by a highly 
efficient algorithm for convergence acceleration [\onlinecite{mixpap}]. The
Brillouin zone integrations were performed using the linear tetrahedron
method with up to 1313 {\bf k}-points within the irreducible wedge of
the rhombohedral Brillouin zone [\onlinecite{aswbook,bloechl94}].

In the present work, a new full-potential version of the
ASW method was employed [\onlinecite{fpasw}]. In this version, the electron
density and related quantities are given by spherical harmonics
expansions inside the muffin-tin spheres. In the remaining
interstitial region, a representation in terms of atom-centered
Hankel functions is used [\onlinecite{msm88}]. However, in contrast to previous
related implementations, we here get away without needing a
so-called multiple-$\kappa$ basis set, which fact allows us to investigate
rather large systems with a minimal effort.

\subsection{Sample preparation and characterization}
\label{sec:spac}

Polycrystalline samples of ${\rm CuCr_{1-x} Mg_x O_2}$, ($0\leq {\rm x} \leq 0.08$)
have been prepared by solid-state reaction by mixing ${\rm Cu_2 O}$, 
${\rm Cr_2 O_3}$ and MgO within the stoichiometric ratios 0.5:0.5(1 -
x):x. The powders were then pressed into bars and fired at 1200 $^o$C for 12
h in air.

As explained in the following, due to the observation of
impurities for low Mg level, to study the possibility of holes created
by a lack of ${\rm Mg^{2+}}$ for ${\rm Cr^{3+}}$ substitution, i.e. vacancies
at the Cu or Cr sites, two samples, ${\rm Cu_{0.98} CrO_2}$ and 
${\rm CuCr_{0.98} O_2}$, were also prepared in the same conditions. A third
sample of formula ${\rm Cu_{1.02} Cr_{0.98} O_2}$ 
was also made to force a Cu substitution for Cr, i.e. to induce
tetravalent chromium if the oxidation of Cu is less than three.
Additionally, as hole creation can result from an oxygen excess,
two samples with different nominal oxygen contents, ${\rm CuCrO_2}$ and
${\rm CuCrO_{2.5}}$, obtained by mixing the precursors ${\rm Cu_2 O}$ or CuO and
${\rm Cr_2 O_3}$, were also prepared by firing, at the same temperature, the
bars in silica tubes sealed under primary vacuum. To also check
for oxygenation non-equilibrium phenomena, these samples were
either cooled by quenching in air from 1200  $^o$C or by cooling at
100 $^o$C/h and 50 $^o$C/h.

The quality of the obtained bars was systematically checked
by X-ray powder diffraction (XRPD) data collected by using a
PANalytical X'pert Pro diffractometer 
(Cu $K_{\alpha}$, $12^o \leq 2\theta \leq 112^o$).
A Zeiss SUPRA 55 scanning electron microscope was also used
to observe the microstructure of the samples. The cationic
compositions were checked by an EDAX energy dispersive X-ray spectroscopy (EDS) system. Transmission electron microscopy
analyses were also carried out with a JEOL 2010CX microscope
equipped with an EDS INCA analyser. It must be emphasized that
the low level of magnesium content to be detected is at the limit
of the analyser. The sample preparation for these observations
was made by crushing in butanol small pieces of bars and the
corresponding microcrystals were deposited on Ni grids. On the
EDS spectrum, in addition to the characteristic peaks of the Cu, Cr
and Mg elements, peaks of Ti, C and Ni were also observed, due to
the sample holder and Ni grid.

The resistivity ($\rho$) measurements were made by the four-probe
technique using ultrasonically deposited indium contacts. The
temperature or magnetic field dependent ρ data were collected by
using a Quantum Design physical properties measurement system
(PPMS) equipped with superconducting coils delivering maximum
fields of 7 T or 9 T with temperatures up to 400 K. A steady-state
technique was used in a similar set-up in order to measure the Seebeck coefficient (S) with a typical gradient of T = 1 K. The $\rho$ and
S data were collected upon cooling from 400 K and 315 K, respectively. An Ulvac-Zem3 set-up was also used to extend the resistivity measurements to higher temperatures (T maximum of 1073 K).
The magnetic susceptibility was obtained by dividing the magnetization (M) by the magnetic field (0.3 T) with M data collected in
zero-field cooling mode by using a SQUID magnetometer (MPMS,
Quantum Design, maximum magnetic field of 5 T).

\section{Results}
\subsection{Electronic structure calculations for $ {\rm CuCrO_2}$ : 
Results and discussion} 

\begin{figure}[t]
\centering
\includegraphics*[width=\columnwidth,clip]{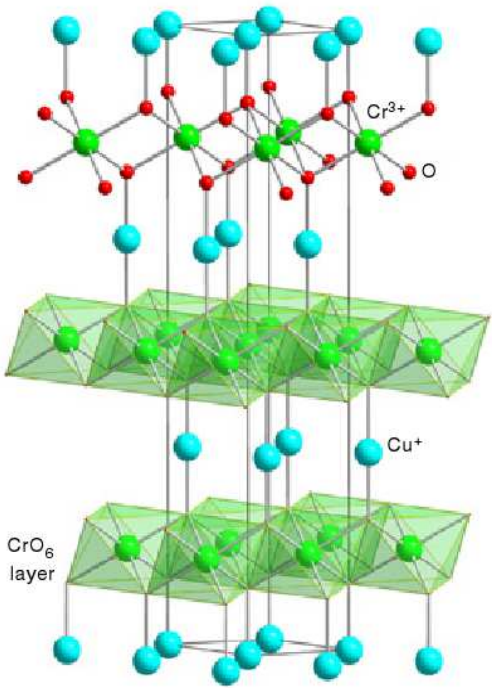}
\caption{Crystal structure of $ {\rm CuCrO_2}$. Copper, chromium and oxygen
  atoms are shown as spheres, blue, green and red, respectively. The
  connection between two successive layers of $ {\rm CrO_6}$ octahedra is made
  through O-Cu-O dumbbells. }
\label{fig:struc}
\end{figure}

The calculations used the crystal structure data reported by
Crottaz et al. [\onlinecite{Crottaz}], and depicted in Fig.~\ref{fig:struc}. 
As a starting point,
spin-degenerate calculations for the rhombohedral structure were
performed. In the resulting partial densities of states (DOS) the
lower part of the spectrum is dominated by O $ 2p $ states and
the transition metal d states lead to rather sharp peaks in the
interval from -4 to +3 eV. In particular, one obtains the $t_{2g}$ and
$e_g$ manifolds of the Cr $ 3d $ states as resulting from the octahedral
coordination. This representation of the partial DOS refers to a
local rotated coordinate system with the Cartesian axes pointing
towards the oxygen atoms. $\sigma$-type overlap of the O $ 2p $ states with
the Cr $ 3d $ $e_g$ orbitals leads to the contribution of the latter between
-7 and -6 eV. In contrast, due to the much weaker $\pi$-type overlap
of the O $ 2p $ states with the $t_{2g}$ orbitals, these states give rise to
sharp peaks in the interval from -1.0 to 0.3 eV. The Fermi energy
falls into the middle of the $t_{2g}$ manifold and gives rise to a 
$ d^3 $ state.
From this and the fact that $E_F$ is very close to the highest peak we
would expect long-range ferromagnetic ordering of Cr moments
of 3 $\mu_B$ in a spin-polarized calculation. This result is compatible
with the scenario used in the cobaltites in which the $t_{2g}$ orbitals
splitted by the rhombohedral distortion are responsible for the
large values of the Seebeck coefficient [\onlinecite{Singh00}]. The Cu 3d
states are 
essentially limited to the interval from -4 to -1 eV and thus Cu
can be assigned a monovalent d10 configuration in close analogy
with the experimental findings.

While previous studies reported a possible $120^o$ magnetic
structure [\onlinecite{Kadowaki}], more recent neutron diffraction experiments
confirm that the magnetic structure is far more complex
[\onlinecite{Poienar}]. In view of this still unsettled situation, we opted to
perform subsequent 
spin-polarized calculations for an assumed ferromagnetic state in
the same manner as in the previous work by Galakhov et al., by
Ong et al., and by ourselves on $ {\rm CuFeO_2}$
[\onlinecite{galakhov97,ong07,cufeo2}]. In contrast to the 
title compound this material does not show long-range antiferromagnetic order in the ideal rhombohedral structure. Yet, as has
been pointed out by Ye et al., $ {\rm CuFeO_2}$ undergoes several structural
distortions and assumes a monoclinic structure with space group
C2/m at 4 K [\onlinecite{Ye}]. The latter allows for an antiferromagnetically
ordered ground state as has been confirmed by our calculations
[\onlinecite{cufeo2}]. While performing this study on $ {\rm CuFeO_2}$, we
were able to show that the electronic states calculated for the
antiferromagnetic ground state and an assumed ferromagnet are very similar apart
from the fact that the former displays a finite optical band gap.
Actually, this striking similarity was just a consequence of the
strongly localized nature of the Fe $ 3d $ $t_{2g}$ states. To conclude,
using an assumed ferromagnetic state for investigating the electronic properties of the title compound is well justified as long as
we are interested in the electronic properties.

\begin{figure}[b]
\centering
\includegraphics[width=\columnwidth,clip]{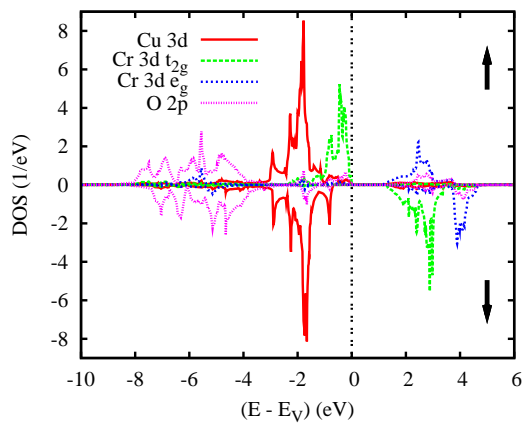}
\caption{Partial densities of states (DOS) of 
         rhombohedral $ {\rm CuCrO_2} $.}
\label{fig2}
\end{figure}

From these calculations, a stable ferromagnetic configuration
was obtained with magnetic moments of 3.0 $\mu_B$. Most importantly,
the density of states as displayed in Fig.~\ref{fig2} reveals the opening of a
fundamental band gap of about 1.2 eV between the spin-up and
spin-down Cr $ 3d $ $t_{2g}$ states, which also carry the overwhelming
part of the magnetic moment. In contrast, the polarization of O
$ 2p $ states is rather low as is expected from the small overlap with
the $t_{2g}$ states (see Ref. [\onlinecite{cufeo2,pdcoo2}] for a more detailed discussion).
In passing, we mention that according to the above-mentioned
results for $ {\rm CuFeO_2}$ we expect very similar electronic states from
spin-polarized calculations using long-range antiferromagnetic
order. In particular, due to its reduced symmetry, such a state is
very likely to also display insulating behaviour.

\begin{figure}[htb]
\centering
\includegraphics[width=0.8\columnwidth]{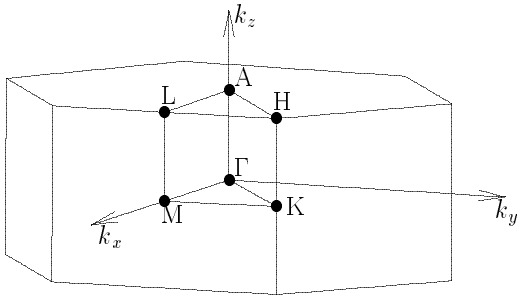}
\caption{First Brillouin zone of the hexagonal lattice.}
\label{fig3}
\end{figure}

\begin{figure}[htb]
\centering
\includegraphics[width=\columnwidth,clip]{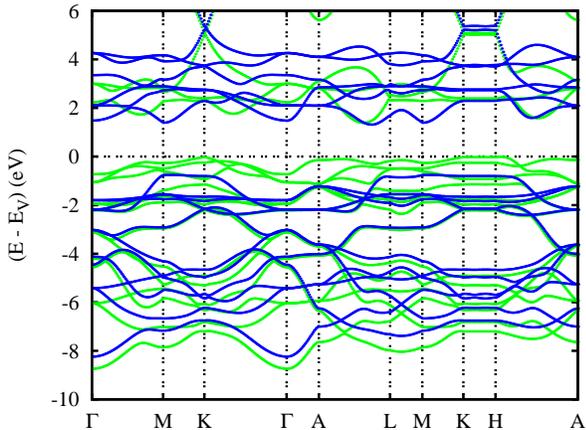}
\caption{Electronic bands of rhombohedral ferromagnetic $ {\rm CuCrO_2}$. 
Green (blue) curves correspond to the majority (minority) spin bands.} 
\label{fig4}
\end{figure}

The electronic bands along selected high-symmetry lines of the
first Brillouin zone of the hexagonal lattice, Fig.~\ref{fig3}, are displayed
in Fig.~\ref{fig4}. 
Clearly their dispersion in the vicinity of the Fermi
energy exhibits an intricate behaviour. First, the dispersion is quite
substantial in most directions, pointing towards a considerable
three-dimensionality of the electronic states arising from the
coupling between the layers as has been observed also in other
delafossite materials [\onlinecite{cufeo2,pdcoo2,singh07}]. In particular, its
magnitudes along $\Gamma-$K and $\Gamma-$A are very similar. Second, 
we obtain that two bands
are reaching the Fermi energy, both at the K and the H point.
Nevertheless their dispersion along K-H (and M-L) is much smaller
than along the $\Gamma-$A direction.

These calculations allow drawing a number of conclusions.
First of all, the dispersion of the highest occupied bands is
remarkably small, as can be seen from the bands around the
H and K points. Accordingly, assuming a vanishingly small hole
doping, one expects a natural tendency towards localization of the
low-energy excitations, even without strong correlations. Besides,
including the latter is expected to enhance this tendency towards
localization. The localized character of these excitations may even
result in a structural transition, as happens in, for example, the
related compound $ {\rm CuFeO_2}$ [\onlinecite{Ye}], where the distorted
structure is further stabilized by superexchange processes
[\onlinecite{cufeo2}]. In $ {\rm CuCrO_2}$, 
the situation appears less extreme, as no structural distortion is
observed. This may be due to the fact that the spin carried by
the $ {\rm Cr^{3+}}$ ions is smaller than the one carried by the $ {\rm Fe^{3+}}$ ions. In
that case the amount of superexchange energy that is lost due to
frustration is reduced, and there is no need for the system to distort
in order to recover that energy.

Finally, in a rigid band picture, one can anticipate that hole
doping should result in peculiar effects. For very small doping, even
though the Fermi velocity increases, it remains small. In such a
case, the formation of small polarons is expected and the system
should remain insulating while mostly keeping the magnetic
structure of the undoped compound. Under further doping the
resulting increase of the Fermi velocity should result in a more
metallic behaviour, accompanied by spin dependent transport.

In order to check these predictions, we performed a series of
susceptibility and transport measurements on weakly Mg-doped
$ {\rm CuCrO_2}$ samples.

\subsection{Structural study}
\label{sec:strs}

\begin{figure}[t]
\centering
\includegraphics*[width=\columnwidth,clip]{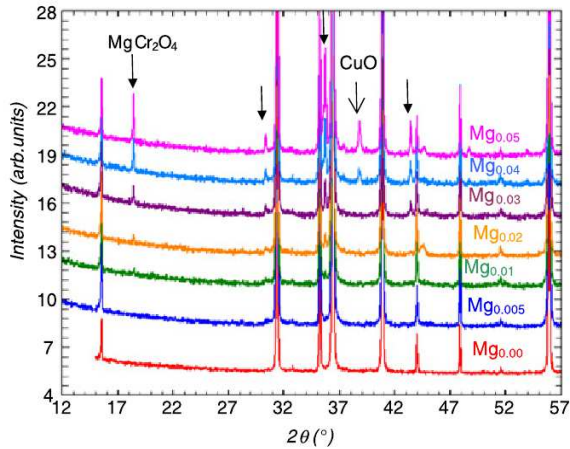}
\caption{Enlargement of the low-angle part of the X-ray diffraction
patterns of the $ {\rm CuCr_{1-x} Mg_x O_2}$ (x = 0.005, 0.01, 0.02, 0.03,
0.04, 0.05) samples 
characteristic of the delafossite phase. $\downarrow$ and $\downarrow$ indicate
peaks attributed to the $ {\rm MgCr_2 O_4}$ spinel and CuO that appear as
impurities.} 
\label{fig:xray}
\end{figure}

\begin{figure}[t]
\centering
\includegraphics*[width=\columnwidth,clip]{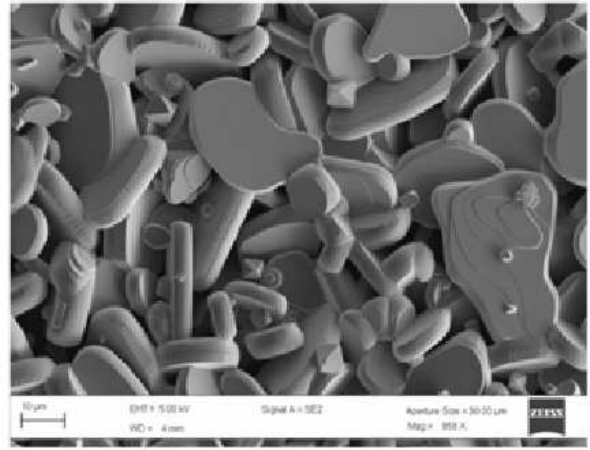}
\caption{SEM image of $ {\rm CuCr_{0.99} Mg_{0.01} O_2}$. The largest platelet-like microcrystals
correspond to the delafossite phase whereas the small crystals of octahedral shape
are attributed to spinel impurity.
}
\label{fig:micro}
\end{figure}

The X-ray powder diffraction study of the $ {\rm CuCr_{1-x} Mg_x O_2}$
compounds prepared in air reveals a very limited $ {\rm Mg^{2+}}$ solubility
range, as shown by the patterns in Fig.~\ref{fig:xray}. In addition to the main
diffraction peaks coming from the 
$R \bar{3} m$ 
delafossite phase (with
$a \simeq 2.97$ and $c \simeq 17.10$~\AA $ $ in the hexagonal setting), extra peaks
(indicated by black arrows in Fig.~\ref{fig:xray}) corresponding to the
$Fd\bar{3}m$ space group with $a \simeq 8.33$~\AA $ $ and thus attributed to 
$ {\rm MgCr_2 O_4}$ can be detected as soon as $ {\rm x} \simeq 0.01$. Furthermore,
the presence of a second impurity phase, identified as CuO, is also observed
for the compounds corresponding to x = 0.04 and 0.05. This preliminary
analysis is consistent with the observations made by scanning
electron microscopy. As shown in Fig.~\ref{fig:micro} for 
$ {\rm CuCr_{0.99} Mg_{0.01} O_2}$, the
material is composed mainly of large plate-like grains (the EDS
analysis of which leads to a Cu/Cr ratio close to 1) and of smaller
grains of octahedral shape, corresponding to the spinel phase (in
agreement with the EDS results: Mg/Cr = 0.5).

At first glance, these results point towards an apparent lack
of $ {\rm Mg^{2+}}$ solubility or a very limited solubility range. However, as
shown in the next section, by RT measurement of the Seebeck
coefficient S, which is a sensitive probe of the charge carrier
concentration, positive $ S $ values for all x are found with a
magnitude that decreases as x increases up to $ {\rm x} \simeq 0.02$. This
indicates an increase of the hole concentration in the nominal
$ {\rm CuCr_{1-x} Mg_x O_2}$ formula. By combining these results 
-- in agreement with previous reports -- 
with the study of the X-ray patterns showing the
presence of impurities as soon as $ {\rm CuCr_{0.99} Mg_{0.01} O_2}$, it
appears that in addition to the Mg substitution, non-stoichiometry
phenomena could also be responsible for creating holes in ${\rm Cu Cr O_2}$.
In order to address this crucial problem, several additional
experiments were made as aforementioned in Section~\ref{sec:spac}. The 
$ S $
values were measured at 300 K for all samples. For all samples
without magnesium (with cation deficiencies or excess oxygen
and also for the sample with Cu/Cr $ > 1 $) prepared at the same
conditions as ${\rm CuCr_{1-x} Mg_x O_2}$ no significant $ S $ change compared
to ${\rm Cu Cr O_2}$ could be evidenced, with all $ S $ values at 300 K lying
near $S \simeq 1 {\rm mV/K}$. These compounds were also post-annealed
under oxygen at 100 atm ($600 ^o$C, 12 h) but without significant
effect on the transport properties. 
Thus our results show that the
oxygen content does not change significantly in those delafossite
samples. This is confirmed by an independent neutron powder
diffraction study of air-prepared polycrystalline samples of ${\rm Cu Cr O_2}$
and ${\rm CuCr_{0.98} Mg_{0.02} O_2}$ (x = 0.02) which confirms their ‘‘O$_2$ ’’
stoichiometry (in the accuracy of the technique) [\onlinecite{Poienar}]. All the
corresponding $S_{\rm 300~K}$ values are close to $\sim 1 {\rm mV/K}$ for the Mg-
free samples. Such a value is similar to the one reported in
Ref. [\onlinecite{Zcucrmg}] but much higher than that of
Ref. [\onlinecite{Okuda}]. This strongly suggests that the samples in
Ref. [\onlinecite{Okuda}] have chemical formulas different from the nominal
compositions. 

Finally, the Mg presence in the delafossite micrograins was
demonstrated by using EDS analysis coupled to electron diffraction
in the transmission electron microscope for the air-prepared
$ {\rm CuCr_{0.98}Mg_{0.02}O_2} $ sample. As shown in 
Fig.~\ref{fig:micro}, the analyzed
regions were chosen on the basis of their electron diffraction
(ED) patterns characteristic of the delafossite structure (c-axis
parameter of $\simeq 17$~\AA $ $ compatible with the $R\bar{3}m$ space group of the
delafossite). The EDS analysis statistics made on 10 microcrystals
leads to an average measured Mg content $ {\rm x_{mes.}} \simeq 0.01$. Although
the proof of the Mg substitution in ${\rm Cu Cr O_2}$ is established, its low
concentration makes very difficult a quantitative determination.
This very limited substitution range of about 1\% to be compared
to the nominal content of 2\% could be explained by the poor
adaptability of the compact ${\rm MO_2}$ layers in this structural type. The
larger ${\rm Mg^{2+}}$ ionic radius than that of ${\rm Cr^{3+}}$, 
$r_{Cr^{3+}}$ = 0.0615 nm against $r_{Mg^{2+}}$ = 0.072 nm, could be the
reason for the limited range of Mg substitution.

\begin{figure}[htb]
\centering
\includegraphics*[width=\columnwidth,clip]{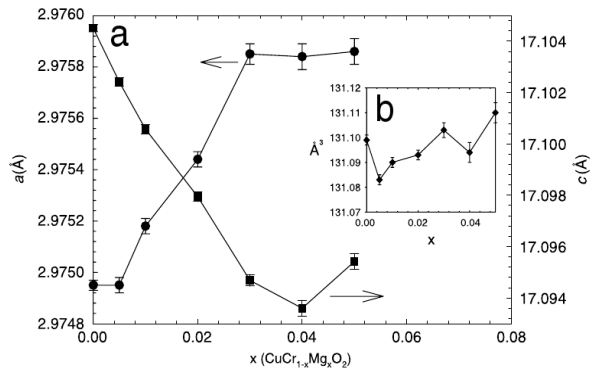}
\caption{Main panel (a) Unit cell parameters (a: circles, left y-axis ; c:
  squares, right y-axis) as a function of the nominal $ {\rm Mg^{2+}}$ content
  x in $ {\rm CuCr_{1-x} Mg_x O_2}$. Inset (b) x dependence of the
  corresponding unit cell volume. }
\label{fig:latpar}
\end{figure}
In order to compare the structural parameters to those reported
in previous studies [\onlinecite{Okuda}], the X-ray data of the present samples
were also refined in the delafossite structure ($R\bar{3}m$, space group).
According to the unit cell parameters and volumes given in
Fig.~\ref{fig:latpar}, 
the changes are small but beyond the error of the technique,
with $(a_0 - a_{0.04} )/a_0 = -0.03\%$ and $(c_0 - c_{0.04} )/c_0 = 0.06\%$,
i.e. clearly smaller than what is reported in Ref. [\onlinecite{Okuda}] (0.3\%
and 0.2\%, respectively). Moreover, plateaus in the $a$ (or $c$) $= f (x)$
curves are observed for ${\rm x} \geq 0.03$ which also differ from the
monotonic evolution reported in Ref. [\onlinecite{Okuda}]. Finally, it is
found that the unit cell volume remains almost constant as x increases (in the
range 131.08 \AA$^3$ -- 131.11 \AA$^3$), which is consistent with the limited
magnesium solubility (inset of Fig. ~\ref{fig:latpar}).

In conclusion of this structural part, it is found that the Mg
substitution is limited to about $\sim 1\%$. This maximum is obtained
as soon as x = 0.02 in ${\rm CuCr_{1-x} Mg_x O_2}$. It must be emphasized
that this very low solubility is lower than previously reported in
Ref. [\onlinecite{Okuda,Zcucrmg}]. This very low amount of substitution also explains the
limited variation of the unit cell parameters with the substitution.
Remarkably, this small amount is sufficient to strongly affect
the transport properties. However, one cannot exclude that this
substitution is also accompanied by other hardly measurable non-stoichiometry
phenomena such as changes in the Cu/Cr ratio 
and/or the oxygen content. But what is clear is the appearance of
the ${\rm MgCr_2 O_4}$ spinel impurity.

\subsection{Physical measurements}\label{sec:phymes}

\begin{figure}[htb]
\centering
\includegraphics*[width=\columnwidth,clip]{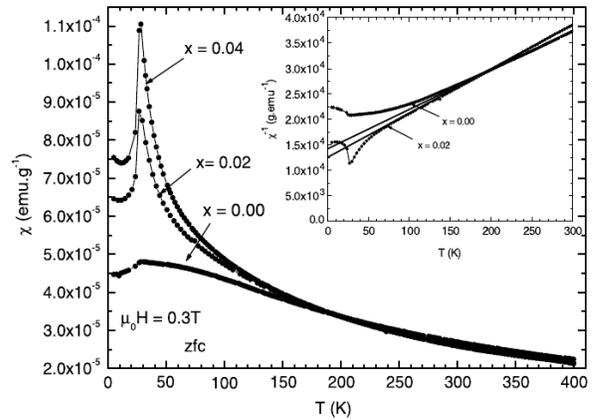}
\caption{T -dependent magnetic susceptibility ($\chi$) for samples of the series
$ {\rm CuCr_{1-x} Mg_x O_2}$ (x values are labelled in the graph). Inset: 
T-dependent reciprocal 
magnetic susceptibility ($\chi^{-1}$) for x = 0.00 and x = 0.02; the straight
lines are for the Curie–Weiss fitting curves.
 }
\label{fig:susc}
\end{figure}

\begin{figure}[b]
\centering
\includegraphics*[width=\columnwidth,clip]{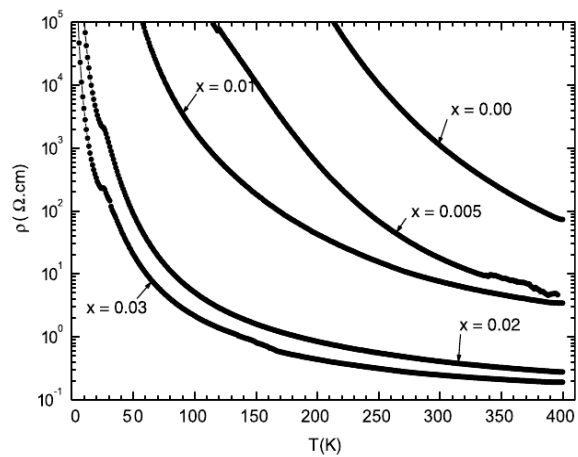}
\caption{T -dependent electrical resistivity ($\rho$) for the series 
$ {\rm CuCr_{1-x} Mg_x O_2}$.
 }
\label{fig:rho}
\end{figure}

\begin{figure}[t]
\centering
\includegraphics*[width=\columnwidth,clip]{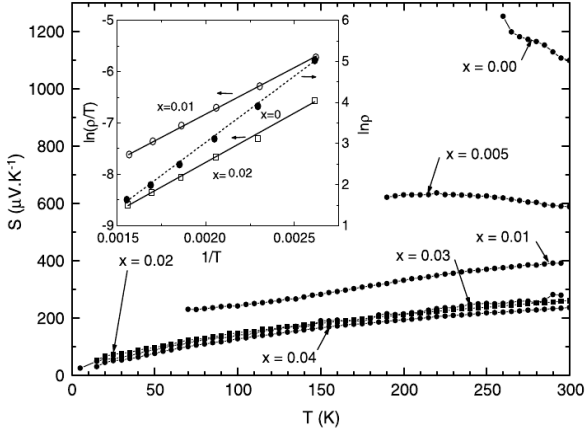}
\caption{T -dependent Seebeck coefficient (S) for the series 
$ {\rm CuCr_{1-x} Mg_x O_2}$. Inset:
inverse temperature dependence of the high temperature resistivity values for
x = 0.00 (right y-scale, $\ln{(\rho)}$), x = 0.01 and x = 0.02 (left y-scale, $\ln{(\rho/T)}$); the
straight lines correspond to the activated behaviour (x = 0.00) and small polaron
fitting (x = 0.01 and x = 0.02).}
\label{fig:rho-high}
\end{figure}

At first glance, it is difficult to relate the structural results
to the strong impact induced by the ${\rm Mg^{2+}}$ substitution on the
magnetic susceptibility $\chi$ (Fig.~\ref{fig:susc}), electrical resistivity
$\rho$ (Fig.~\ref{fig:rho}) and Seebeck coefficient $ S $ 
(Fig.~\ref{fig:rho-high}).  
First, it must be pointed out that
the ${\rm CuCrO_2}$ sample with $\rho_{\rm 300~K}\ =\ 1\ {\rm k\Omega~cm}$, 
$S_{\rm 300~K}\ =\ 1.1\ {\rm mV/K}$ appears to be less self-doped than in the
previous study of Ref. [\onlinecite{Okuda}] ($\rho_{\rm 300~K}\ =\ 0.2\ 
{\rm k\Omega~cm}$, 
$S_{\rm 300~K}\ =\ 0.35\ {\rm mV/K}$). Second, the $\chi (T)$ curve of 
${\rm CuCrO_2}$ reveals a characteristic downturn below $\sim 24\ {\rm K}$ 
(Fig.~\ref{fig:susc}) which is in good agreement with the value
$T_N$ = 24 K reported in the literature for ${\rm CuCrO_2}$
[\onlinecite{Okuda,Kadowaki,Poienar}]. More 
importantly, the $\chi$ values just above $T_N$ are smaller than those
of the ${\rm Mg^{2+}}$ substituted ${\rm CuCrO_2}$ compounds
(Fig.~\ref{fig:susc}). This is also
clearly seen as well-defined $\chi (T)$ peaks at $T_N \sim 25\ {\rm K}$ observed
for the doped compounds compared to a smoother maximum for
${\rm CuCrO_2}$. However, the magnetic structure determined by neutron
diffraction data refinements is not dramatically changed by the 2\%
Mg substitution [\onlinecite{Poienar}]. The analysis of the $\chi (T)$ curve of ${\rm CuCrO_2}$
and ${\rm CuCr_{0.98} Mg_{0.02} O_2}$ (inset of Fig.~\ref{fig:susc}), by using
the Curie–Weiss law ($\chi = \frac{C}{T - \theta_{CW}})$, indicates for the
former a narrow T-range for the $\chi^{-1}$ linear regime. Both the upward
deviation of the $\chi^{-1}(T)$ curve starting below about 200 K and the
Curie–Weiss constant value, $\theta_{CW} \sim -170\ {\rm K}$, indicate large
antiferromagnetic fluctuations 
in the undoped compound. However, the effective paramagnetic
moment ($\mu_{\rm eff}$) values extracted from the linear part of the
$\chi^{-1}(T)$ curves, $\mu_{\rm eff} \simeq 3.7-4.0 \mu_B$, do not
significantly change with x. 
This value is close to the spin-only value for high spin ${\rm Cr^{3+}}$ with
S = 3/2, leading to 2[S (S + 1)]1/2 = 3.87\ $\mu_B$. These values
and the trend with x are in good agreement with those previously
reported [\onlinecite{Okuda}] and also with the ${\rm Cr^{3+}}$ high spin state coming from
the band structure calculations.

\begin{figure}[t]
\centering
\includegraphics*[width=\columnwidth,clip]{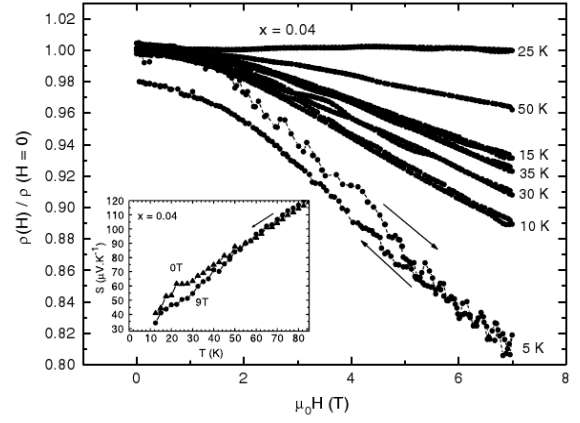}
\caption{Isothermal magnetic field dependent resistivity of $ {\rm CuCr_{0.96}
    Mg_{0.04} O_2}$. Inset: low temperature enlargement of the $S(T)$ curves
  for $ {\rm CuCr_{0.96} Mg_{0.04} O_2}$ collected upon cooling from 315K in 0T and then in 9T.}
\label{fig:mrho}
\end{figure}

Interestingly, the impact of the ${\rm Mg^{2+}}$ doping on the magnetic
properties is accompanied by a dramatic change of the electronic
properties. A large decrease of the electrical resistivity $\rho$ values is
induced, $\rho$ decreasing at 300~K from $1\ {\rm k\Omega~cm}$ in ${\rm CuCrO_2}$
down to $0.3\ {\rm \Omega~cm}$ in $ {\rm CuCr_{0.99}Mg_{0.01}O_2} $
(Fig.~\ref{fig:rho}). Moreover, the Mg doping 
also affects the temperature dependence of the resistivity. This is
illustrated by the curves shown in the inset of Fig.~\ref{fig:rho-high} showing
the $\rho$ data collected for $T > \rm 300~K$, i.e. far beyond the T region
where the magnetic fluctuations might play a role. Indeed, for the
undoped compound, the temperature dependence of the resistivity
(in the inset of Fig.~\ref{fig:rho-high}) is following an activated behaviour,
$\ln{\rho} \propto T^{-1}$, over the temperature range where it is
measurable, with an activation energy of 280 meV. In contrast, this does not
hold 
true for the doped compound, where a small polaron behaviour,
$\ln{(\rho/T)} \propto T^{-1}$, is observed in the same range of temperature
with an activation energy of 170-190 meV for x = 0.01 and x = 0.02 in
the inset of Fig.~\ref{fig:rho-high}, as anticipated above. 
By further
increasing the nominal content of ${\rm Mg^{2+}}$, the $\rho(T)$ curves keep
roughly the same 
shape and values in agreement with the limited solubility of ${\rm Mg^{2+}}$
reached for ${\rm x} \geq 0.02$. As already reported in Ref. [\onlinecite{Okuda}], a kink is
found at $T_N$ for the doped samples (Fig.~\ref{fig:rho}), suggesting a
spin-charge coupling. This is confirmed in the case of the conducting samples
by the existence of a negative magnetoresistance (MR) (Fig.~\ref{fig:mrho}),
for ${\rm CuCr_{0.96} Mg_{0.04} O_2}$. At 5 K, this effect reaches $ -19 $\% 
in 7 T)
[\% MR = 100 $\times (\frac{\rho(H)-\rho(H =0)}{\rho(H=0)})$]. However, as T
is increased, the 
isothermal $\rho(H)/\rho(H=0)$ curves show that the MR magnitude
follows a non-monotonic behaviour: as T increases from 5 K, the
MR first decreases (curves collected at 10, 15, 25 K in Fig.~\ref{fig:mrho}),
so that MR $\simeq$ 0 at $T_N =\ 25\ {\rm K}$; then MR increases again just
above $T_N$ (T = 30 K), and finally progressively decreases (35 K
and 50 K curves). For this last T region ($T > T_N$), MR is observed
up to T $\sim$ 70 K (see also the [$\rho(H=0)/\rho(H=7\ T)$]$(T)$
curve for ${\rm CuCr_{0.96} Mg_{0.04} O_2}$ in the inset of
Fig.~\ref{fig:mrho}). Thus, these compounds behave as if the application of a
magnetic field above $T_N$ reduces the spin scattering of charges leading to
negative MR. This strongly resembles the conventional effect observed of
spin-polarized transport in ferromagnetic conducting oxides beyond $T_C$. But
as the magnetic ordered state is antiferromagnetic, at $T_N$, 
the setting of the long range antiferromagnetism blocks charge
hopping with the corresponding MR = 0. The following negative
MR for $T < T_N$ is again due to the re-entrance of the regime due to
spin-polarized tranport.

The observed negative magnetoresistance strongly supports
that the ${\rm Cr^{3+}}$/${\rm Cr^{4+}}$  mixed valence in the ${\rm CrO_2}$
layer rather than ${\rm Cu^+}$/${\rm Cu^{2+}}$ is responsible for the
observed electrical conductivity. 
In that respect, the analysis of the Seebeck coefficient values
gives important information. As shown in Fig.~\ref{fig:rho-high}, the
$S_{\rm 300~K}$ values decrease from $1.1\ {\rm mV/K}$ for ${\rm CuCrO_2}$ 
to $0.39\ {\rm mV/K}$ for ${\rm CuCr_{0.99} Mg_{0.01} O_2}$, and then 
$0.26\ {\rm mV/K}$ for ${\rm CuCr_{0.98} Mg_{0.02} O_2}$. The
latter value is in good agreement with the data reported by Ono
et al. [\onlinecite{Zcucrmg}]. For higher x values in ${\rm CuCr_{1-x} Mg_x O_2}$, $S_{\rm 300~K}$ 
saturates to about $0.25-0.27\ {\rm mV/K}$. This is consistent with the
limited ${\rm Mg^{2+}}$ solubility of 1\% reached for nominal magnesium
contents such as ${\rm x} \geq 0.02$ in ${\rm CuCr_{1-x} MgO_2}$. This explains the
rather constant charge carrier concentration for ${\rm x} > 0.01$.

This $ S $ decrease as x increases can be simply explained by using
the Heikes formula, $S\ =\ \frac{k_B}{e} [\ln{(\frac{1-x}{x})}]$, where x
corresponds to the ${\rm Cr^{4+}}$ density. A good agreement is found for 
${\rm CuCr_{0.96} Mg_{0.01} O_2}$ : $S_{\rm Heikes}\ =\ 396\ \mu {\rm V/K}$ 
against $S_{\rm exp} = 390 \mu {\rm V/K}$. Moreover, for ${\rm CuCrO_2}$, 
$S_{\rm exp}\  = 1.1\ {\rm mV/K}$ results into 
${\rm x}_{\rm calc}\ =\ 2.9\ 10^{-6}$, i.e. a very small fraction of 
${\rm Cr^{4+}}$. For larger x values, the presence of
impurities revealed by X-ray diffraction precludes the analysis of
the data. It must be emphasized that a charge compensation to the
${\rm Mg^{2+}}$ substitution by creating ${\rm Cu^{2+}}$ would give the same
result as a fraction of x(${\rm Mg^{2+}}$) is creating x(${\rm Cu^{2+}}$) 
corresponding to holes in a ${\rm Cu^+}$ matrix. In order to rule out one of
these hypotheses, Seebeck measurements have been measured without and within
magnetic field as shown for ${\rm CuCr_{0.96} Mg_{0.04} O_2}$ in the inset of
Fig.~\ref{fig:mrho}. 

Applying an external magnetic field upon cooling makes the
$ S $ values decrease compared to the $S_{H=0} (T)$ measurements. This
negative magnetothermopower, which reaches a maximum at $T_N$,
indicates a contribution of the spins to the thermopower pointing
towards a spin-polarized transport in the ${\rm CrO_2}$ planes. A similar
effect was found in cobaltites with isostructural $ {\rm CoO_2} $ layers
[\onlinecite{Limelette}]. 
The existence of concomitant negative magnetoresistance and
magnetothermopower in these delafossites, both related to $T_N$,
thus supports an electronic conduction by ‘‘${\rm Cr^{4+}}$’’ holes in the 
${\rm CrO_2}$ magnetic layer.

The most striking result of the present study is the strong
impact on the magnetic and transport properties induced by
low ${\rm Mg^{2+}}$ doping. It must be recalled that the ${\rm Mg^{2+}}$
solubility 
appears to be very limited as attested by the impurity formation
already detected for the nominal composition ${\rm CuCr_{0.99} Mg_{0.01} O_2}$.
In agreement, a maximum of 1\% Mg can be measured, as
shown from the study of the compound of nominal composition
${\rm CuCr_{0.98} Mg_{0.02} O_2}$. This explains the very weak x-dependence of
$\rho$, $ S $ and $\chi$ observed for ${\rm x} \geq 0.02$ in ${\rm CuCr_{1-x} Mg_x O_2}$.

\section{Discussion and concluding remarks}
The present study reveals that the ${\rm Mg^{2+}}$ solubility in the
${\rm CuCr_{1-x} Mg_x O_2}$ delafossite is limited to ∼1\%. As shown by the
electronic band structure, in the ${\rm CuCrO_2}$ starting material, the main 
contributions, at the valence band maximum, are given by the
${\rm Cr^{3+}}$ t$_{2g}$ orbitals, the Cu $ 3d $ states lying much deeper in
energy. As a result, in the absence of doping, the compound is
semiconducting. This is consistent with the large Seebeck coefficient at room 
temperature, $S\ = 1.1\ {\rm mV/K}$. In contrast to ${\rm PdCoO_2}$, for
which the lack of ${\rm Co^{3+}}$ contribution of the ${\rm CoO_2}$ layer to
the electronic states gives a strong 2D character to these states mainly
driven by the Pd$^+$ species, the ${\rm CrO_2}$ contribution creates 3D
electronic states ensuring a strong interlayer coupling. The latter is
attested by the long range non-commensurate antiferromagnetism of 
${\rm CuCrO_2}$ [\onlinecite{Poienar}]. The spin-polarized calculations point towards a band gap
between spin-up 
and spin-down with $3 \mu_B$ per ${\rm Cr^{3+}}$. The discrepancies between our
$\rho$ and $ S $ data and those of previous studies for the starting ${\rm CuCrO_2}$
oxide reflect its great sensitivity to doping. This study demonstrates that
complete sets of synthesis and structural characterizations are needed to
exclude the presence of non-stoichiometry 
phenomena that could be very tiny. For instance, the larger $ S $ values
at 300~K, indicating smaller hole concentration, point towards the
more stoichiometric character of our samples as compared to those
of Ref. [\onlinecite{Okuda}]. In addition, the presence of impurities, detected even
for the x = 0.01 compound, shows that the measurements made
in previous studies for samples containing probably some impurities could not
easily be trusted. As far as electrical transport is 
concerned, these impurities could act as parallel circuits, thus rendering the
interpretations difficult. 

Starting from the electronic structure of ${\rm CuCrO_2}$, hole doping
(‘‘${\rm Cr^{4+}}$’’) in the ${\rm Cr^{3+}}$ matrix would create charge
carrier delocalisation in a spin-polarized media. This scenario based on a
transport  made on the Cr–O network rather than on the Cu one is confirmed
by the bump observed at $T_N$ on the $\rho(T)$ curves for ${\rm x} \geq 0.02$, the
negative magnetothermopower and magnetoresistance. For the
latter, the spin scattering reduction induced by the application of
a magnetic field is responsible for the negative MR observed above
$T_N$. At $T_N$, the antiferromagnetic setting blocks the spin scattering
so that magnetic field application has no longer an effect on the
electrical resistivity (in Fig.~\ref{fig:mrho} and within 7 T, MR = 0 at 
$T_N\ = 25\ {\rm K}$ to be compared to $ {\rm MR = - 9} $\% at 30 K). 
In that respect, the
re-entrance of the MR for $T < T_N$ might correspond to remaining disordered
spin regions which locally allow some magnetic moment reorientation under the
application of a magnetic field. 

One of the major results of this study is that only 1\% ${\rm Mg^{2+}}$
substitution for ${\rm Cr^{3+}}$, formally creating 1\% of ${\rm Cr^{4+}}$, is
sufficient to drastically modify the electronic properties. Such an effect
can be compared to the doping effects in transparent conducting
oxides as ${\rm Sn^{4+}}$-doped ${\rm In_2 O_3}$ (ITO). However, the magnetism
of the high spin states of chromium trivalent and tetravalent cations in 
${\rm CuCrO_2}$ is responsible for spin polarization. Accordingly, the drastic
decrease of the electrical resistivity induced by the substitution is
reinforcing the magnetic ordering via the coupling between holes
and spins. This explains the higher $T_N$ value observed for 
Mg-substituted ${\rm CuCrO_2}$ [\onlinecite{Okuda,Poienar}].

\section{Acknowledgments}
We gratefully acknowledge many useful discussions with T.\ Kopp, V.\ Hardy,
Ch.\ Simon and W.\ C.\ Sheets.
This work was supported by the Deutsche Forschungsgemeinschaft through
SFB 484.


\begin{thebibliography}{}
\bibitem{Bednorz} 
J.G. Bednorz, K.A. M\"uller, Z. Phys. B {\bf 64}, 189 (1986).

\bibitem{Mitzushima} 
K. Mitzushima, et al., Mat. Res. Bull. {\bf 15}, 783 (1980).

\bibitem{Terasaki} 
I. Terasaki, Y. Sasago, K. Uchinokura, Phys. Rev. B {\bf 56},  R 12685 (1997).

\bibitem{Takada}
K. Takada, et al., Nature {\bf 422}, 53 (2003).

\bibitem{Jansen}
M. Von Jansen, R. Hoppe, Z. Anorg, Allg. Chem. {\bf 408}, 104 (1974).

\bibitem{Singh00} 
D.J. Singh, Phys. Rev. B {\bf 61}, 13397 (2000).

\bibitem{Takeda}
K. Takeda, et al., J. Phys. Soc. Japan {\bf 63}, 2017 (1994).

\bibitem{Kimura06}
T.\ Kimura, J.\ C.\ Lashley, and A.\ P.\ Ramirez, 
Phys.\ Rev.\ B {\bf 73}, 220401(R) (2006).

\bibitem{Pabst}
A. Pabst, Am. Mineral. {\bf 23}, 175 (1938).

\bibitem{Ye}
F. Ye, et al., Phys. Rev. B {\bf 73 }, 220404 (2006).

\bibitem{Okuda}
T. Okuda, et al., Phys. Rev. B {\bf 72}, 144403 (2005);
T. Okuda, et al., Phys. Rev. B {\bf 77}, 134423 (2008).

\bibitem{Zcucrmg} 
Y.\ Ono, K.\ Satoh, T.\ Nozaki, and T.\ Kajitani,
Jap.\ J.\ Appl.\ Phys.\ {\bf 46}, 1071 (2007).

\bibitem{Kadowaki}
H. Kadowaki, H. Kikuchi, Y. Ajiro, J. Phys.: Condens. Matter {\bf 2}, 4485 (1990).

\bibitem{Poienar}
M. Poienar, et al., Phys. Rev. B {\bf 79}, 014412 (2009).

\bibitem{Seki08} 
S.\ Seki, Y.\ Onose, and Y.\ Tokura,
Phys.\ Rev.\ Lett.\ {\bf 101}, 067204 (2008). 

\bibitem{Perdew96}
J.P. Perdew, K. Burke, M. Ernzerhof, Phys. Rev. Lett. {\bf 77}, 3865 (1996).

\bibitem{Perdew92}
J.P. Perdew, Y. Wang, Phys. Rev. B {\bf 45}, 13244 (1992).

\bibitem{wkg}
A.\ R.\ Williams, J.\ K\"ubler, and C.\ D.\ Gelatt, Jr., 
Phys.\ Rev.\ B {\bf 19}, 6094 (1979).

\bibitem{aswrev}
V.\ Eyert, 
Int.\ J.\ Quantum Chem.\, {\bf 77}, 1007 (2000).

\bibitem{aswbook}
V.\ Eyert,
{\em The Augmented Spherical Wave Method -- A Comprehensive Treatment},
Lect.\ Notes Phys.\ {\bf 719} (Springer, Berlin Heidelberg 2007).

\bibitem{sgo}
V.\ Eyert and K.-H.\ H\"ock, 
Phys.\ Rev.\ B {\bf 57}, 12727 (1998).

\bibitem{mixpap}
V.\ Eyert, 
J.\ Comp.\ Phys.\ {\bf 124}, 271 (1996).

\bibitem{bloechl94}
P.\ E.\ Bl\"ochl, O.\ Jepsen, and O.\ K.\ Andersen,
Phys.\ Rev.\ B {\bf 49}, 16223 (1994).

\bibitem{fpasw}
V.\ Eyert, 
unpublished 

\bibitem{msm88}
M.\ S.\ Methfessel, 
Phys.\ Rev.\ B {\bf 38}, 1537 (1988).

\bibitem{Crottaz}
O. Crottaz, F. Kubel, H. Schmid, J. Solid State Chem. {\bf 122}, 247 (1996).

\bibitem{galakhov97}
V.\ R.\ Galakhov, A.\ I.\ Poteryaev, E.\ Z.\ Kurmaev, V.\ I.\ Anisimov,
S.\ Bartkowski, M.\ Neumann, Z.\ W.\ Lu, B.\ M.\ Klein, and T.-R.\ Zhao,
Phys.\ Rev.\ B {\bf 56}, 4584 (1997).

\bibitem{ong07}
K.\ P.\ Ong, K.\ Bai, P.\ Blaha, and P.\ Wu,
Chem.\ Mater.\ {\bf 19}, 634 (2007).

\bibitem{cufeo2}
V.\ Eyert, R.\ Fr\'esard, and A.\ Maignan,
Phys.\ Rev.\ B {\bf 78}, 052402 (2008).

\bibitem{pdcoo2}
V.\ Eyert, R.\ Fr\'esard, and A.\ Maignan,
Chem.\ Mat.\ {\bf 20}, 2370 (2008).

\bibitem{singh07}
D.\ J.\ Singh,
Phys.\ Rev.\ B {\bf 76}, 085110 (2007).

\bibitem{Limelette} 
P. Limelette, et al., Phys. Rev. Lett. {\bf 97}, 046601  (2006).

\end{thebibliography}
\end{document}